\documentclass{aastex}

\shorttitle{On the Ages of Resonant Planetary Systems}
\shortauthors{Koriski \& Zucker}

\begin{document}

\title{On the Ages of Planetary Systems with Mean Motion Resonances}
\author{Shuki Koriski and Shay Zucker}
\affil{Dept. of Geophysics \& Planetary Sciences, 
	Raymond and Beverly Sackler Faculty of Exact Sciences, Tel
	Aviv University, Tel Aviv 69978, Israel}
\email{jehoshu1@post.tau.ac.il,	shayz@post.tau.ac.il}

\begin{abstract}

We present preliminary though statistically significant evidence that
shows that multiplanetary systems that exhibit a $2/1$ period
commensurability are in general younger than multiplanetary systems
without commensurabilities, or even systems with other
commenurabilities. An immediate possible conclusion is that the $2/1$
mean-motion resonance in planetary systems, tends to be disrupted
after typically a few Gyrs.

\end{abstract}


\section{INTRODUCTION}
\label{intro}

The number of known extrasolar multiplanetary systems is constantly
growing. A significant fraction of these systems exhibit what seems to
be couples of planets that are locked in mean-motion resonances
(MMR). The question whether a specific system is indeed in a state of
MMR is usually hard to answer conclusively. Period commensurability
(PC) is a necessary condition for MMR, but not sufficient. Full
diagnosis of an MMR requires long-term monitoring of the resonant
argument, which is impractical in many cases. Nevertheless,
\citet{Wrietal2011a} used PC as a proxy indicator for MMR and showed,
in a brilliantly simple statistical test, that planetary systems
exhibit PCs much more than expected in random. Since PC is not known
to relate to any other physical phenomenon, except for MMR, the simple
conclusion is that MMR is a preferred state of planetary systems
\citep[e.g.,][]{Beaetal2008}.

Models show that planets can get trapped in a state of MMR through
convergent orbital migration \cite[e.g.,][]{PapSzu2010,Ketetal2011} or
planet-planet scattering \cite[e.g.,][]{Rayetal2008}.  The long-term
survivability of MMR is also the focus of several studies. Chaotic
diffusion may lead eventually to disruption of the resonance
\cite[e.g.,][]{TisMal2009,GozMig2009}. \citet{Thoetal2008} claim that
the presence of a remnant planetesimal disk may be a common reason for
the disruption of such resonances.

It seems that the most prominent MMR is the $2/1$
resonance. \citet{PieNel2008} show that the $2/1$ resonance is a
preferred outcome of the orbital evolution of two planets embedded in
a protoplanetary disc. \citet{Micetal2008a,Micetal2008b} draw a
detailed portrait of the phase space of the $2/1$ resonance, while
\citet{Voyetal2009} provide a detailed analysis of the families of
solutions to the general dynamical problem of three bodies in $2/1$
MMR, addressing specifically the issue of stability.

If indeed most MMRs are destined to get destabilized and disrupted, as
\citet{Thoetal2008} claim, we would expect to see a shortage of MMRs
around older stars. This is the phenomenological claim we set out to
examine in this short study. The results we obtained were a little
different -- we found out that only stars hosting planets in a $2/1$
PC were statistically significantly younger. All other PCs do not seem
to show such an age relation.

In the next section we describe the way we built the sample on which
we tested our hypothesis, Section \ref{tests} details the statistical
tests we applied on this sample to test our claims, and we conclude in
Section \ref{discussion}.

\section{THE SAMPLE}
\label{sample}

We first built a sample of stars hosting known multiplanet systems,
using the publicly available exoplanet orbit database that
\citet{Wrietal2011b} have put up online. In order to make our sample as
homogeneous as possible, we considered only planets that were detected
by radial velocities, around stars of spectral types F, G, or K. Thus,
we excluded most of the known transiting planets (except those
detected first through radial velocities), pulsar planets, planets
detected in direct imaging, planets around M stars, and the Solar
System planets. 

Next, we identified those systems that exhibited period
commensurabilities. We included in our definition of
commensurabilities integer ratios larger than $1$, with a denominator
less than $6$, i.e., the ratios $2/1$, $3/1$, $3/2$, $4/1$, $4/3$,
$5/1$, $5/2$, $5/3$, and $5/4$. In order to tag two periods as
commensurate we defined a 'normalized commensurability proximity'
(NCP) score, defined by: $$ \delta=2 \frac{|r-r_c|}{r+r_c} $$ where
$r$ is the actually measured period ratio and $r_c$ is the
commensurability ratio against which we compare $r$. We chose to tag
systems with at least one NCP value of $\delta<0.1$ as commensurate.

Next we had to introduce an age estimate for the stars in our
sample. Stellar ages are notoriously difficult to
estimate. \citet{Sod2010} reviewed and compared several age estimation
approaches. There are two approaches that dominate the literature. The
first uses the stellar activity, as estimated by the H and K lines of
singly ionized Calcium in the stellar spectrum. The second places a
star on model isochrones on the Hertzsprung-Russell diagram. Both
methods, as well as the less frequently used methods, are strongly
model dependent and suffer many drawbacks and pitfalls.

For the sake of sample homogeneity, we decided to focus purely on one
approach.  Furthermore, Fig.~8 in the paper by \citet{Sod2010} shows
that besides a prevailing systematic shift between isochrone ages and
chromospheric activity ages, it seems that isochrone ages might lose
their sensitivity for stars younger than about
$2\,\mathrm{Gyr}$. Thus, we decided to use in our study only
chromospheric activiy ages based on the Calcium H and K emission
lines.  To avoid non uniformities in the interpretation of
observations, we extracted the chromospheric activity ages only from
large surveys we found in the literature, and not from papers that
presented analysis of individual stars. Our use of chromospheric
activity ages is also another reason for excluding M stars from our
sample, as M stars are notorious for having a variable activity
\citep{Sod2010}.

Table~\ref{resonant_systems} presents the resulting sample of
commensurate planetary systems, including the relevant
commensurability ratios and the NCP values. The table also lists the
chromospheric activity ages we found in the literature, and the
average age we computed from these
values. Table~\ref{non_resonant_systems} lists the systems that did
not pass our criterion for PC ($\delta<0.1$) and their relevant
ages. In both tables we included also systems for which we did not
find any chromospheric activity age in any large published survey. In
total, our sample conveniently includes $15$ age estimates for
commensurate systems, and $15$ for non-commensurate systems.

\begin{deluxetable}{rrlllllc}
\tabletypesize{\scriptsize}
\rotate
\tablecaption{The sample of commensurate multiplanetary systems \label{resonant_systems}}
\tablewidth{0pt}
\tablehead{
\colhead{HD} &
\colhead{HIP} &
\colhead{Other} &
\colhead{Commensurabilities} &
\colhead{Normalized Commensurability} &
\colhead{Published Ages} &
\colhead{Reference} &
\colhead{Mean Age} \\
\colhead{} &
\colhead{} &
\colhead{Names} &
\colhead{} &
\colhead{Proximity} &
\colhead{(Gyr)} &
\colhead{} &
\colhead{(Gyr)} }
\startdata
9826 & 7513 & $\upsilon$ And & $5/1$ & $0.058$ & $6.31$,$5.32$,$6.23$,$7.26$ & 1,2,3,4 &$6.28$ \\
10180 & 7599 & \nodata & $4/1$,$5/1$,$5/2$,$3/1$,$3/1$ & $0.087$,$0.022$,$0.014$,$0.014$,$0.055$ & $6.46$ & 5 & $6.46$ \\
37124 & 26381 & GJ 209 & $2/1$ & $0.048$ & $3.89$,$3.33$,$4.72$ & 1,2,3 & 3.98 \\
40307 & 27887 & GJ 2046 & $2/1$,$5/1$ & $0.061$,$0.052$ & \nodata & \nodata & \nodata \\
45364 & 30579 & \nodata & $3/2$ & $0.007$ & $4.87$ & 5 & $4.87$ \\
60532 & 36795 & GJ 279 & $3/1$ & $0.000$ & \nodata & \nodata & \nodata \\
69830 & 40693 & GJ 302 & $4/1$ & $0.094$ & $4.68$,$6.36$,$6.43$,$6.1$ & 1,3,4,6 & $5.89$ \\
73526 & 42282 & \nodata & $2/1$ & $0.004$ & $5.59$ & 2 &$5.59$ \\
75732 & 43587 & 55 Cnc & $3/1$ & $0.010$ & $6.46$,$5.5$,$6.44$,$3.43$,$8.7$ & 1,2,3,5,6 & $6.11$ \\
82943 & 47007 & \nodata & $2/1$ & $0.002$ & $4.07$,$3.08$,$5.10$ & 1,2,3 & $4.08$ \\
\nodata & \nodata & BD +20 2457 & $5/3$ & $0.021$ & \nodata & \nodata & \nodata \\
90043 & 50887 & 24 Sex & $2/1$ & $0.029$ & \nodata & \nodata & \nodata \\
108874 & 61028 & \nodata & $4/1$ & $0.063$ & $7.41$,$7.26$ & 1,2 & $7.33$ \\
115617 & 64924 & 61 Vir & $3/1$ & $0.075$ & $6.31$,$5.90$,$6.62$,$6.1$ & 1,3,4,6 & $6.23$ \\
128311 & 71395 & GJ 3860 & $2/1$ & $0.017$ & $0.39$,$0.43$ & 2,4 & $0.41$ \\
155358 & 83949 & \nodata & $5/2$ & $0.084$ & $5.32$ & 3 & $5.32$ \\
160691 & 86796 & $\mu$ Ara & $2/1$ & $0.035$ & $6.41$,$3.31$,$6.5$ & 2,5,6 & $5.41$ \\
181433 & 95467 & GJ 756.1 & $5/2$ & $0.087$ & \nodata & \nodata & \nodata \\
183263 & 95740 & \nodata & $5/1$ & $0.011$ & $8.13$,$7.38$ & 1,3 & $7.75$ \\
200964 & 104202 & \nodata & $4/3$ & $0.008$ & \nodata & \nodata & \nodata \\
202206 & 104903 & \nodata & $5/1$ & $0.076$ & $2.04$,$2.95$ & 2,7 & $2.49$
\enddata
\tablerefs{
(1) \cite{Wrietal2004}
(2) \cite{Safetal2005}
(3) \cite{IsaFis2010}
(4) \cite{Maletal2010}
(5) \cite{RocMac1998}
(6) \cite{MamHil2008}
(7) \cite{Arr2011} }
\end{deluxetable}

\begin{deluxetable}{rrlllc}
\tabletypesize{\footnotesize}
\tablecaption{The sample of non-commensurate multiplanetary systems \label{non_resonant_systems}}
\tablewidth{0pt}
\tablehead{
\colhead{HD} &
\colhead{HIP} &
\colhead{Other} &
\colhead{Published Ages} &
\colhead{Reference} & 
\colhead{Mean Age} \\
\colhead{} &
\colhead{} &
\colhead{Names} &
\colhead{(Gyr)} &
\colhead{} &
\colhead{(Gyr)} }
\startdata
9446 & 7245 & \nodata & \nodata & \nodata & \nodata \\
11964 & 9094 & GJ 81.1A & $9.55$,$9.56$ & 1,2 & $9.55$ \\
12661 & 9683 & \nodata & $7.41$,$7.05$,$7.05$ & 1,2,3 & $7.17$ \\
\nodata & 14810 & \nodata &  $7.77$ & 3 & $7.77$ \\
38529 & 27253 & \nodata & $4.90$,$5.09$,$6.73$ & 1,2,3 & $5.57$ \\
47186 & 31540 & \nodata & $2.72$,$8.13$ & 5,7& $5.43$ \\
74156 & 42723 & \nodata & $7.24$,$7.38$,$7.54$ & 1,2,3 & $7.39$ \\
\nodata & 40967 & BD -08 2823 & \nodata & \nodata & \nodata \\
95128 & 53721 & 47 UMa, GJ 407 & $6.03$,$6.03$,$6.10$,$4.93$,$4.4$ & 1,2,3,4,6 & $5.50$ \\
125612 & 70123 & \nodata & $4.23$ & 3 & $4.23$ \\
134987 & 74500 & 23 Lib, GJ 579.4 & $7.76$,$7.32$,$8.12$ & 1,2,3 & $7.73$ \\
147018 & 80250 & \nodata & $3.16$ & 7 & $3.16$ \\
168443 & 89844 & GJ 4052 & $8.51$,$5.90$,$8.19$ & 1,2,3 & $7.53$ \\
169830 & 90485 & \nodata & $7.24$,$4.95$ & 1,2 & $6.09$ \\
187123 & 97336 & \nodata & $6.31$,$5.33$,$6.59$ & 1,2,3 & $6.08$ \\
190360 & 98767 & GJ 777A & $7.76$,$7.09$,$8.6$ & 1,2,6 & $7.82$ \\
215497 & 112441 & \nodata & \nodata &\nodata & \nodata \\
217107 & 113421 & \nodata & $7.41$,$7.32$,$8.19$ & 1,2,3 & $7.64$
\enddata
\tablecomments{
For references, see Table~\ref{resonant_systems}. }
\end{deluxetable}

\section{STATISTICAL TESTS}
\label{tests}

The mean chromospheric activity age of the commensurate systems in our
sample is $5.213\,\mathrm{Gyr}$, while that of the non-commensurate
systems is $6.577\,\mathrm{Gyr}$. This difference of
$1.36\,\mathrm{Gyr}$ hints that resonant systems tend to be younger on
average. In order to test this hypothesis, we adopted the most simple
approach of the permutation test \citep{Goo1994}. Thus, we repeatedly
drew a random assigment of the ages to the two samples, effectively
ruining any correlation that may exist between age and
commensurability. For each such random assignment we recalculated the
mean age difference. We used $10^6$ random assignments, among which
$20650$ yielded an age difference larger than
$1.36\,\mathrm{Gyr}$. This implies a statistical significance of
$p=0.021$.

The main advantage of the permutation test approach is in avoiding the
need to assume any special assumptions about the distribution of the
samples. However, one may still argue that using the mean values is
prone to a strong influence by the extreme values in each sample. An
alternative is to use the median instead, which is more robust to
extreme values. The median age of the resonant systems is
$5.59\,\mathrm{Gyr}$, and that of the non-resonant systems
$7.17\,\mathrm{Gyr}$, with a difference of $1.58\,\mathrm{Gyr}$. We
repeated the permutation test, this time obtaining $17178$ out of
$10^6$ values larger than the actual value. Thus, the permutation test
for the medians leads to a somewhat more significant result, with a
significance of $p=0.017$.

The results we have presented above are only marginally
significant. They do seem to point to a tendency of the commensurate
systems to be younger than the non-commensurate ones, but their
statistical significance is not that high. Further examination of the
sample shows that the tendency we see may be attributed only to the
$2/1$ PC systems.  Close examination of Table~\ref{resonant_systems}
hints that the subsset of the $2/1$ PC systems (HD\,37124, HD\,73526,
HD\,82943, HD\,128311, and $\mu$\,Ara) seem to possess lower ages, The
additional two $2/1$ PC systems HD\,40307 and 24\,Sex do not have an
age estimate, and thus do not contribute to the statistical
significance. Since the number of $2/1$ PC systems is much smaller
than the total number of PC systems, it is not immediately obvious
that this result is statistically significant. We repeated the tests
we performed earlier, this time dividing the sample into $2/1$ PC
systems, and all the rest. This new division clearly enhances the
statistical significance: The 'difference in means' test now yields
$p=0.007$ ($6745$ out of $10^6$), and the 'difference in medians' test
gives $p=0.004$ ($4178$ out of $10^6$). The actual age difference is
$2.40\,\mathrm{Gyr}$ for the difference of the means, and
$2.15\,\mathrm{Gyr}$ for the difference of the medians.

Further tests we have performed showed that the rest of the
commensurate systems did not exhibit any significant age difference
compared to the non-commensurate systems.

\section{CONCLUSION}
\label{discussion}

The results we presented in this short Letter support the claim that
the phenomenon of mean-motion resonance, which manifests itself as
period commensurability, is not generally ever lasting. The actual
numbers we obtained suggest that a typical life expectancy of a $2/1$
mean-motion resonance is around $4\,\mathrm{Gyr}$. For the other
families of resonances we cannot assert at this stage any
statistically significant claim, probably because no other category is
as pupulated yet as the $2/1$ category.

The above conclusion is extremely simplistic. It does not take into
account the details of the orbits involved in the resonance, such as
mass ratios or eccentricities. It is also prone to large and
significant uncertainties, which are known to plague stellar age
estimates. However, the scarcity of the current dataset does not allow
for a more detailed and refined analysis.

Our results suggest that the $2/1$ resonance stands out among all the
resonances. This may very well be the case, as \citet{PieNel2008} have
claimed. It might be that the orbital evolutionary history of the
$2/1$ resonance is unique and different from that of all the other
resonances. In fact, \cite{PieNel2008} also singled out the $3/2$
resonance as another preferred outcome of the resonance trapping
scenario. Our analysis may support this, as the only $3/2$ PC system
in the sample (HD\,45364) is indeed younger than average
($4.87\,\mathrm{Gyr}$). Since there is currently only one system in
this category, we chose not to include this claim in our tests, even
though it would have surely improved the statistical significance.

In order to explain the scarcity of $2/1$ PC among the older systems,
one needs to invoke some mechanism to disrupt them. Thus, our results
agree with the claim by \cite{Thoetal2008} that breakup (maybe
violent) of resonantly-locked planets is a common evolutionary step of
planetary systems. The fact that non-$2/1$ resonances seem to survive,
may hint that their formation is an outcome of a much later stage in
the evolution of planetary systems. In order to test this possibility,
it is essential to perform much more long-term dynamical studies of
resonant systems, lasting a few Gyrs and more.

In order to further explore the issue of survivability of mean-motion
resonances, we need also to refine our knowledge of multiplanetary
systems. Specifically, we should compile a more comprehensive dataset
of stellar ages for the multiplanetary systems. Hopefully, with the
advent of the recent planet finding missions, such data will become
more abundant.

The results we presented in this Letter are only a preliminary attempt
to test whether the issue of survival of mean-motion resonsnces is
worth exploring with the tools of stellar age estimates. Apparently,
the existing data partly corroborate the hypothesis we presented in
Section~\ref{intro}, and the $2/1$ PC indeed tends to be found in
younger systems. This may very well be another window into the
understanding of planetary orbital evolution.

\acknowledgments
This research was supported by the ISRAEL SCIENCE FOUNDATION -- The
Adler Foundation for Space Research (grant No. 119/07).  This research
has made use of the Exoplanet Orbit Database and the Exoplanet Data
Explorer at exoplanets.org.

\end{document}